# Symmetries and invariant solutions of the wave equation for shear disturbances in soft solids

Alexander I. Kozlov[a]

*Chair of medical and biological physics,
Vitebsk State Medical University, Republic of Belarus,
210023*

The Lie-group approach was applied to determine symmetries of the third-order nonlinear equation formulated for description of shear elastic disturbances in soft solids. Invariant solutions to that equation are derived and it turned out that they could represent outgoing or incoming exponentially decaying or unbounded disturbances.

## I. INTRODUCTION

Soft solids (like gels or some biological tissues) differ from Newtonian liquids in particular with possibility to maintain shear stresses and therefore with ability to guide transverse elastic waves. Nonlinear wave equation for shear elastic disturbances was considered in last decades in many works (see [1-3], for example). Nonlinear elastic constants as well as viscosity were taken in account so that the partial differential equation of the third order in derivatives was proposed for the one-dimentional case [2-3]:

$$\frac{1}{c^2}\frac{\partial^2 u}{\partial t^2} = \frac{\partial^2 u}{\partial z^2} + \frac{2}{3}\beta\frac{\partial}{\partial z}\left(\frac{\partial u}{\partial z}\right)^3 + \tau\frac{\partial}{\partial z}\left\{\left[1 + 2\left(\frac{\partial u}{\partial z}\right)^2\right]\frac{\partial^2 u}{\partial t \partial z}\right\} \quad (1)$$

where $z$ and $t$ are the spatial and time coordinates respectively, $c$ is the small-signal transverse-sound-wave velocity, $\tau$ is proportional to coefficients of viscosity and is defined in [2] as $\tau = (\zeta + 4\eta/3)/\mu$ ($\zeta$ and $\eta$ are so-called the second and the first viscosity coefficients [4]), $\mu$ is the second of the Lamé's elastic constants, $\beta = 3(\mu + A/2 + D)/(2\mu)$ is the nonlinearity coefficient, where $A$ and $D$ are the third-order and the forth-order elastic constants [1-3].

Different solutions of Equation (1) were investigated analytically and numerically [1-3], and an attempt to derive one-way equation of the lower order in derivatives has been made [3]. Nevertheless it seems to be useful to investigate symmetries of the partial differential equation (1), because this approach can bring some new exact analytical solutions for problems, which were solved only approximately or numerically for years [5].

This short communication is devoted to application of methods of the classical theory of Lie groups for solution of Equation (1).

---

[a] Electronic addresses for correspondence: albapasserby@yahoo.com or aikozlovvgmu@gmail.com.

## II. SOLUTION

Before solving Eq. (1) its variables were changed in the next way [3]:

$$w = \sqrt{\frac{2}{3}\beta} \cdot u \qquad \theta = \frac{t}{\tau} \qquad x = \frac{z}{c\tau} \qquad \alpha = \frac{3}{c^2 \tau^2}$$

So we obtain the equation

$$w_{\theta\theta} = w_{xx} + \alpha w_x^2 w_{xx} + w_{\theta xx} + 2\frac{\alpha}{\beta} w_x w_{\theta x} w_{xx} + \frac{\alpha}{\beta} w_x^2 w_{\theta xx} \qquad (2)$$

where lower indices denote differentiation with respect to appropriate coordinates, as usual.
Appling the standard Lie-group approach we consider the next operator of one-parametric group of infinitesimal transformation, which is prolonged to all necessary derivatives [6-7]:

$$X = \xi^\theta \frac{\partial}{\partial \theta} + \xi^x \frac{\partial}{\partial x} + \eta^w \frac{\partial}{\partial w} + \zeta^\theta \frac{\partial}{\partial w_\theta} + \zeta^x \frac{\partial}{\partial w_x} + \zeta^{\theta\theta} \frac{\partial}{\partial w_{\theta\theta}} + \zeta^{xx} \frac{\partial}{\partial w_{xx}} \zeta^{\theta x} \frac{\partial}{\partial \eta_{\theta x}} + \zeta^{\theta xx} \frac{\partial}{\partial \eta_{\theta xx}}$$

Applying the latter differential operator to Equation (2) we can get the determining equations of the problem:

$$\xi^\theta_\theta = \xi^\theta_x = \xi^\theta_w = 0 \qquad \xi^x_\theta = \xi^x_x = \xi^x_w = 0 \qquad \eta^w_{\theta\theta} = \eta^w_x = \eta^w_w = 0$$

Solution of these PDE gives the next operators of infinitesimal transformation of Equation (2):

$$X_1 = \frac{\partial}{\partial \theta} \qquad X_2 = \frac{\partial}{\partial x} \qquad X_3 = (1 + 2\theta)\frac{\partial}{\partial w}$$

Linear combination of these three operators (where $Q$ and $R$ are some arbitrary constants)

$$X = Q\frac{\partial}{\partial \theta} + R\frac{\partial}{\partial x} + (M + 2N\theta)\frac{\partial}{\partial w}$$

leads to the next system of characteristic PDE (where $M$ and $N$ are arbitrary constants) :

$$\frac{d\theta}{Q} = \frac{dx}{R} = \frac{dw}{M + 2N\theta}$$

Solving the latter system we obtain two invariants of Eq. (2): the independent variable $\lambda$ and depending on it $\Phi(\lambda)$, obeying the next expressions:

$$\lambda = Qx - R\theta \quad \text{and} \quad \Phi(\lambda) = Qw - M\theta - N\theta^2$$

Finding $w$ from the second of Equations (3) and substituting the result in (2) we can reduce Equation (2) to the next ordinary differential equation in function $\Phi(\lambda)$ (with primes denoting differentiation):

$$\left(\frac{R^2}{Q}-Q\right)\Phi''_{\lambda\lambda}+\frac{2N}{Q}-\alpha Q(\Phi'_\lambda)^2\Phi''_{\lambda\lambda}+RQ\Phi'''_{\lambda\lambda\lambda}+2\frac{\alpha}{\beta}RQ\Phi'_\lambda(\Phi''_{\lambda\lambda})^2+\frac{\alpha}{\beta}RQ(\Phi'_\lambda)^2\Phi'''_{\lambda\lambda\lambda}=0$$

This ODE can be easily integrated once, the result looks like

$$\frac{R^2}{Q}\Phi'_\lambda+\frac{2N}{Q}\lambda-Q\Phi'_\lambda-\frac{\alpha Q}{3}(\Phi'_\lambda)^3+RQ\Phi''_{\lambda\lambda}+\frac{\alpha}{\beta}RQ(\Phi'_\lambda)^2\Phi''_{\lambda\lambda}=C_1 \qquad (3)$$

where $C_1$ is the constant of integration. If $C_1=N=0$ then a hope remains that the final solution could be not surely unbounded and Equation (3) in this case is reduced to the homogeneous ODE in $\Phi'(\lambda)$:

$$\frac{R^2}{Q}\Phi'_\lambda-Q\Phi'_\lambda-\frac{\alpha Q}{3}(\Phi'_\lambda)^3+RQ\Phi''_{\lambda\lambda}+\frac{\alpha}{\beta}RQ(\Phi'_\lambda)^2\Phi''_{\lambda\lambda}=0 \qquad (4)$$

If we suppose

$$\frac{R}{Q}=\pm\sqrt{1-\frac{\beta}{3}} \qquad (5)$$

then the general solution of Equation (4) obeys the next explicit formula:

$$\Phi(\lambda)=\frac{3R}{\beta}C_2\exp\left(\frac{\beta\lambda}{3R}\right)+C_3$$

where $C_2$ and $C_3$ are other constants of integration. In this case for the general solution to Equation (2) we have

$$w=\frac{3}{\beta}C_2\sqrt{1-\frac{\beta}{3}}\exp\left(\frac{\beta\lambda}{3R}\right)+C_3+\frac{M}{Q}\theta \qquad (6)$$

on condition (5). So the appropriate solution to the initial Equation (1) could be found to be
a) in the particular case $R=Q=1$ (outgoing disturbance)

$$u=\pm C_2\sqrt{\frac{3}{\beta}\left(1-\frac{\beta}{3}\right)}\exp\left(\pm\frac{\beta}{3c\tau}\cdot\frac{z-ct}{\sqrt{1-\frac{\beta}{3}}}\right)+C_3+Mt \qquad (7a)$$

and
b) for $R = -Q = -1$ (incoming disturbance)

$$u = \pm C_2 \sqrt{\frac{3}{\beta}\left(1 - \frac{\beta}{3}\right)} \exp\left(\pm \frac{\beta}{3c\tau} \cdot \frac{z+ct}{\sqrt{1-\frac{\beta}{3}}}\right) + C_3 + Mt \qquad (7b)$$

If the constant $M$ is equal to zero these solutions are decaying or unbounded in time and space for $\beta/3 < 1$. Indeed, it had been found experimentally that for some gels [8-9] in the nonlinearity constant

$$\beta = 3(\mu + A/2 + D)/(2\mu)$$

$\mu$ is positive, while $A$ is often negative and some times more than $\mu$. But in some other substances [9] both constants $\mu$ as well as $A$ could be positive, thus leading to the possibility of periodically oscillating in time decaying in space displacements as the expression under the square root in (7) is negative leading to imaginary values of the exponential function. (Unfortunately, the author has not found in literature numerical values of the forth-order elastic constant $D$, believing that its absolute value is less than those of $\mu$ and $A$.)

To make the picture complete, it should be also mentioned the particular solution (incoming and outgoing) of Equation (6), though it represents the unbounded displacement:

$$\Phi(\lambda) = \sqrt{\frac{3}{\alpha}(R^2 - Q^2)}\frac{\lambda}{Q} + C_4 \qquad (8)$$

thus leading to the next trivial solutions to the Equation (1):

$$u = (z \pm ct)\sqrt{\pm \frac{\beta}{\alpha}} + C_4 + Mt \qquad (9)$$

$C_4$ is another one constant of integration.

### III. CONCLUSION

Classical Lie-group analysis of the nonlinear wave equation for soft tissues was conducted and the invariant solutions to that equation were derived. Those solutions turned out to be in fact outgoing and incoming disturbances, which could be unbounded or exponentially decaying in amplitude at infinite time or space coordinate.

### PUBLISHING DATA SHARING POLICY

The data that supports the findings of this study are available within the article.